\documentclass[12pt]{article}
\usepackage{amsmath,epsfig,bm,a4wide}
\newcommand{\vhi}{\varphi}
\newcommand{\bphi}{{\bm \vhi}}
\begin{document}
%
%
\title{\begin{flushright} \mbox{\normalsize IFJPAN-IV-2006-7} \end{flushright}
Propagation of Uncertainty in a Parton Shower
\footnote{This work is partly supported by the EU grant mTkd-CT-2004-510126 
in partnership with the CERN Physics Department and by the Polish Ministry of 
Scientific Research and Information Technology grant No 620/E-77/6.PRUE/DIE 188/2005-2008.}}

\author{Philip Stephens and Andr\'e van Hameren\\
\it Institute of Nuclear Physics, Polish Academy of Sciences\\
\it ul.\ Radzikowskiego, 31-342, Krak\'ow, Poland}
\maketitle
\abstract{
Presented here is a technique of propagating uncertainties through the parton 
shower by means of an alternate event weight. This technique provides a 
mechanism to systematically quantify the effect of variations of certain 
components of the parton shower leading to a novel approach to 
probing the physics implemented in a parton shower code 
and understanding its limitations. Further, this approach can be applied to a 
large class of parton shower algorithms and requires no changes to the 
underlying implementation.
}
%

\section{Introduction}
As we enter a new era of particle physics, precise knowledge of quantum 
chromodynamics (QCD) will become increasingly important in order to understand
the physics beyond the standard model. Currently, one of the most useful tools
for studying QCD is the parton shower approximation. This tool provides a 
mechanism to connect few-parton states to the real world of high-multiplicity 
hadronic final states while retaining the enhanced collinear and soft 
contributions to all orders. 

Use of parton shower Monte Carlos (MC) has become common-place. Often, when 
one needs an estimate of the uncertainty of a MC prediction several different 
MC programs are used and the differences between them is considered the 
error~\cite{Grunewald:2000ju}. Though this technique of estimating the error 
of the MC is generally acceptable, it does little to provide insight into the 
physics. It has been shown~\cite{Gieseke:2004tc} that the uncertainties in 
both the perturbative expansion and the parton distribution functions indeed
can lead to effects of the order of ten percent. We 
propose here a technique in which the known uncertainties of the physics can 
be propagated through the parton shower framework. This technique provides 
alternate weights to an event generated by a MC without having to change the 
basic structure of the MC program. We feel this technique could be valuable 
when determining how various improvements in the parton shower will impact the 
MC predictions. Furthermore, this gives a more satisfactory description of the
errors in a MC prediction.

We begin by applying the variational technique to the parton shower 
probability densities. Using this technique we are able to define the 
appropriate weights associated with the variations. We then define a MC 
implementation and present numerical examples of applying this technique to 
the parton shower MC. We present the algorithm and formula for a variation to 
the running coupling and the structure of the kernel to show the method works 
as expected. We discuss a way that this procedure may be able to be used to 
estimate the effects of next-to-leading log (NLL) terms, additionally we 
consider how to use this technique to map between two parton shower 
implementations which rely on different interpretations of the kinematics.

\section{Variation of Parton Shower}
In many parton 
showers~\cite{Gieseke:2003hm,Sjostrand:2000wi,Jadach:2005bf,Gleisberg:2003xi} one starts 
with the fundamental probability density (for one emission) defined as
\begin{equation}
{\cal P} = f_R(\vec{y}) \exp \left(- \int^{\xi(\vec{y})} d^n \!\vec{\,y}' 
f_V(\!\vec{\,y}')  \right). 
\label{eqn:probdens}
\end{equation}
Here the function $f_R(\vec{y})$ is the distribution of the real emission
while $f_V(\vec{y})$ is the virtual contribution. In both cases the precise
definition of $\vec{y}$ is specific to the implementation. Furthermore, the
limits of integration in the virtual component are also specific to the 
implementation: how the infra-red limit is treated, the definition of
resolvable versus unresolvable emissions and the ordering of variables.

For a time-like shower $f_R = f_V$ and is given by
\begin{equation}
f_R(\vec{y}) = \frac{\alpha_S(g(\vec{y}))}{2 \pi} P(\vec{y}),\label{eqn:psf}
\end{equation}
where $g(\vec{y})$ is some abstract function used to determine the scale of the
running coupling. We find a similar result for the constrained MC~\cite{Jadach:2005bf}; 
for a space-like shower using the backward evolution algorithm we find
$f_R = f_V f(x,\vec{y})$ and
\begin{equation}
f_V(\vec{y}; x) = \frac{\alpha_S(g(\vec{y}))}{2 \pi} \frac{f(x/z,\vec{y})}
{f(x,\vec{y})} P(\vec{y}),
\end{equation}
where $f(x,\vec{y})$ is the PDF at energy fraction $x$ and scale given by 
some combination of the components of $\vec{y}$. We can explicitly see that one
of the components of $\vec{y}$ is $z$, a momentum fraction. 

In the forward (time-like) evolution algorithm, as well as the non-Markovian 
algorithm, $P(\vec{y})$ is just the Alteralli-Parisi~\cite{Altarelli:1977zs} 
splitting function divided by the scale. In the numerical results here we 
consider only the forward evolution algorithm.%

We now define a functional to represent our functions $f_R(\vec{y})$ and
$f_V(\vec{y})$
\begin{equation}
F_R[{\bm \varphi}(\vec{y})] = f_R(\vec{y}) \, ; \, 
F_V[{\bm \varphi}(\vec{y})] = f_V(\vec{y}).
\end{equation}
Here ${\bm \varphi}(\vec{y})$ are the functional components of $F_{R/V}$ which
we want to vary (e.g.\ the running coupling or the kernel). This defines the
distribution of one branching as
\begin{equation}
{\cal P}[{\bm \vhi}(\vec{y})] = F_R[{\bm \vhi}(\vec{y})] \exp \left(
- \int^{\xi(\vec{y})} d^n \!\vec{\,y}' F_V[{\bm \vhi}(\!\vec{\,y}')] \right).
\end{equation}
We can find the variation of this by
\begin{equation}
\delta {\cal P} = {\cal P}[({\bm \vhi} + \delta {\bm \vhi})(\vec{y})] -
{\cal P}[{\bm \vhi}(\vec{y})].
\end{equation}
If we define
\begin{equation}
\delta F_{R/V} = F_{R/V}[({\bm \vhi} + \delta {\bm \vhi})(\vec{y})] -
F_{R/V}[{\bm \vhi}(\vec{y})],
\end{equation}
then
\begin{equation}
\delta {\cal P} = {\cal P} \left( 1 + \frac{\delta F_R}{F_R} \right) \exp
\left( - \int^{\xi(\vec{y})} d^n \!\vec{\,y}' \delta F_V \right) - {\cal P},
\end{equation}
from which we have a weight
\begin{equation}
w \equiv \frac{{\cal P} + \delta {\cal P}}{\cal P} =
\left( 1 + \frac{\delta F_R}{F_R} \right) \exp \left( - \int^{\xi(\vec{y})} d^n
\!\vec{\,y}' \delta F_V \right). \label{eqn:weight}
\end{equation}

If, at first, we assume that $F_{R/V}$ is proportional to a linear product of
functions, $\vhi_a$, and we consider variations of only one function we
find
\begin{equation}
\delta F_{a R/V}(\vec{y}) = \partial_a F_{R/V} [{\bm \varphi}(\vec{y})] \delta
\varphi_a (\vec{y}), \label{eqn:varf}
\end{equation}
with
\begin{equation}
\partial_a F[{\bm \varphi}(\vec{y})] = \frac{\partial F[{\bm \varphi}]}
{\partial \varphi_a}.
\end{equation}

We now turn to varying multiple functions at the same time; again assuming 
that each $F_{R/V}$ is proportional to the linear product of all $\vhi_a$. At 
lowest order in variations 
\begin{equation}
\delta F_{R/V} = \sum_a \frac{\partial_a F_R[{\bm \vhi}(\vec{y})]}{F_R[{\bm
\vhi}(\vec{y})]} \delta \vhi_a(\vec{y}),
\end{equation}
thus the weight is given by
\begin{equation}
w = \left(1 +  \sum_a \frac{\partial_a F_R[{\bm
\vhi}(\vec{y})]}{F_R[{\bm \vhi}(\vec{y})]} \delta \vhi_a(\vec{y}) \right) \exp
\left( - \sum_a \int^{\vec{y}} d^n \!\vec{\,y}' \partial_a F_V[{\bm \vhi}
(\!\vec{\,y}')] \delta \vhi_a(\vec{y}) \right). \label{eqn:mweight}
\end{equation}
One could keep higher order terms of the variations, but the general formula
have not been presented here. The weight for such results, given in 
eqn.~(\ref{eqn:weight}), would be the same with the modified definitions of
$\delta F_{R/V}$.

The weights defined in eqn.~(\ref{eqn:weight}) are relative 
to the original probability density for one emission. To get the total weight 
for the full event, we must consider
\begin{equation}
{\cal P}_E[{\bm \vhi}, \left\{ \vec{y}_i \right\}] = \prod_i {\cal P}[{\bm
\vhi}(\vec{y}_i)], 
\end{equation}
and thus
\begin{equation}
\delta {\cal P}_E = {\cal P}_E[{\bm \vhi} + \delta {\bm \vhi}, \left\{ 
\vec{y}_i \right\}] - {\cal P}_E[{\bm \vhi}, \left\{ \vec{y}_i \right\}].
\end{equation}
This leads to a total event weight given by
\begin{equation}
w_E \equiv \frac{{\cal P}_E + \delta {\cal P}_E}{{\cal P}_E} = \prod_i w_i.
\end{equation}

\section{Example Parton Shower Kinematics}
\begin{figure}[t]
\centering
\epsfig{file=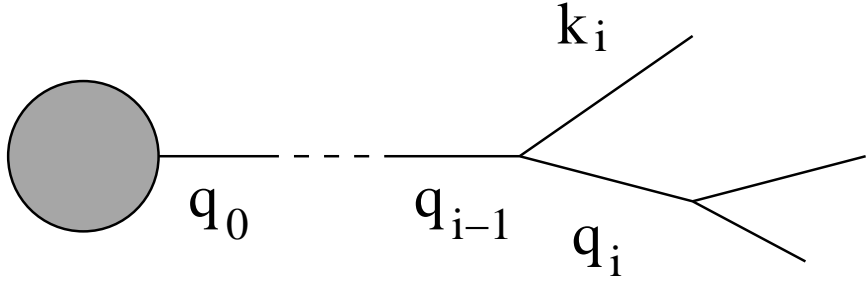}
\caption{Final-state parton branching. The blob represents the hard 
sub-process.\label{fig:brems}}
\end{figure}

For the examples given here we will use as a model bremstrahlung emissions from
one quark line. This is shown in fig.~\ref{fig:brems}. For the numerical 
results presented in the following sections we now define a concrete 
implementation of the kinematics of the parton shower. We will use the 
variables of Herwig++~\cite{Gieseke:2003hm,Gieseke:2003rz}.

We begin by looking at the $i$th gluon emission $q_{i-1} \to q_i + k_i$. In 
the Sudakov basis this is
\begin{equation}
q_i = \alpha_i p + \beta_i n + q_{\perp i},
\end{equation}
with $p$ the jet's ``parent parton'' momentum and $n$ a light-like ``backward''
4-vector. These obey
\begin{equation}
p^2 = m^2 \, ; \, n^2 = 0 \, ; \, p\cdot n = 1 \, ; \, q_{\perp i} \cdot p
= q_{\perp i} \cdot n = 0.
\end{equation}
We find
\begin{equation}
\beta_i = \frac{q_i^2 - q_\perp^2 - \alpha_i^2 m^2}{2 \alpha_i p \cdot n}.
\end{equation}
Lastly we define the momentum fraction and relative transverse momentum as
\begin{equation}
z_i = \frac{\alpha_i}{\alpha_{i-1}}, \, \, {\bm p}_{\perp i} = 
{\bm q}_{\perp i} - z_i {\bm q}_{\perp i-1}.
\end{equation}
One then finds
\begin{equation}
q_{i-1}^2 = \frac{q_i^2}{z_i} + \frac{k_i^2}{1-z_i} + \frac{{\bm p}_{\perp i}^2}{z_i (1-z_i)}.
\end{equation}

The evolution variable used for these examples is
\begin{equation}
\tilde{q}^2 = \frac{\bm{p}_\perp^2}{z^2 (1-z)^2} + \frac{\mu^2}{z^2} + 
\frac{Q_g^2}{z(1-z)^2},\label{eqn:qtilde}
\end{equation}
where $\mu = \max(m,Q_g)$ and $Q_g$ is a cutoff parameter of the model.

Here the splitting kernel, in the quasi-collinear approximation, is
\begin{equation}
P_{qq}(z, \bm{p}_{\perp}^2) = C_F \left[ \frac{1+z^2}{1-z} - 
\frac{2z(1-z)m^2}{\bm{p}_{\perp}^2 + (1-z)^2 m^2}\right]. \label{eqn:kernel}
\end{equation}
With the definition eqn.~(\ref{eqn:qtilde}) the branching probability is
\begin{equation}
dB(q \to qg) = \frac{C_F}{2 \pi} \alpha_S[z^2(1-z)^2 \tilde{q}^2] 
\frac{d \tilde{q}^2}{\tilde{q}^2} \frac{dz}{1-z} \left[ 1+z^2 - 
\frac{2m^2}{z \tilde{q}^2} \right]. 
\end{equation}

The phase-space boundaries are defined by requiring a real transverse
momentum, this is found from eqn.~(\ref{eqn:qtilde}). We denote the solutions
as $z^\pm(\tilde{q}^2)$.

Using the branching probability we can now define the probability density of
our parton shower. Here we find
\begin{equation}
{\cal P} = dB(q\to qg) \exp\left\{-\int dB(q \to qg)\right\},
\end{equation}
which gives
\begin{eqnarray}
F_R[{\bm \vhi}(\vec{y})] &=& 
\frac{C_F}{2\pi} \alpha_S[z^2 (1-z)^2 \tilde{q}^2] \frac{1}{\tilde{q}^2} 
\left[ \frac{1+ z^2}{1-z} - \frac{2m^2}{z(1-z)\tilde{q}^2} \right], \nonumber\\
&\equiv& F[(\alpha_S,P_{qq})(z, \tilde{q}^2)] , \nonumber \\
&=& \frac{1}{2\pi \tilde{q}^2} \alpha_S(z,\tilde{q}^2) 
P_{qq}(z,\tilde{q}^2).
\end{eqnarray}
\section{Running Coupling Uncertainty}
The first variation we wish to study is that of the running coupling. The
running coupling has several sources of uncertainty. For example, one may wish
to study how the variation of the argument to the coupling changes the
results of the MC. This is often done as a practical way to control the 
perturbative series of the running coupling. 

To illustrate the method described in the previous section, we choose the
variation of the coupling to be due to the uncertainty in the measurement of 
the coupling. Standard practice in MCs is to take
the central value of the coupling at $m_Z^2$ and use the two-loop
renormalization group equation (RGE) to run the coupling to alternate scales. 
At the Landau pole, the perturbative series breaks down and one imposes, by 
hand, some treatment of the coupling below that scale. We propose to trace the
uncertainty in the value of the running coupling at $m_Z^2$ through the RGE
and to study the effect of this variation on the predictions of the MC.

Qualitatively, one would expect to see that if we take the upper bound of the
running coupling value that we have more emissions and overall a harder 
$p_\perp$ spectrum of the outgoing quark. If the lower bound is taken, we 
expect the opposite. The magnitude of the fluctuations are governed by the 
evolution of this uncertainty through the RGE. This is exactly what we will 
see. 

We want to stress that the method described in the previous section is
not applicatble only to this particular choice 
of running coupling variance. One could use any choice for the uncertainty of 
the running coupling and the mechanism of propagating this through the parton
shower is unchanged, another option is
\begin{equation}
\delta \alpha_S^+(Q^2) = \alpha_S(Q^2) - \alpha_S(2 Q^2) \, \, ; \, \,
\delta \alpha_S^-(Q^2) = \alpha_S(Q^2) - \alpha_S(Q^2/2),
\end{equation}
for example if one wanted to estimate the error due to the truncation of the
perturbative series for $\alpha_S$.

In most modern applications one uses the two-loop RGE for the running coupling.
For illustrative purposes in the following examples we choose the running coupling to 
take the value $\alpha_S(m_Z^2) = 0.117 \pm 10\%$. We then use 
the running defined by two-loop RGE to run to all scales. In practice we solve
for $\Lambda^2_{QCD}$ which gives the correct value of $\alpha_S(m_Z^2)$ and 
find the upper and lower variances of $\Lambda^2_{QCD}$ to give the 10\% 
variation. These are then used as input to give the running coupling at 
alternate scales; $m_Z$ is chosen as $91.2~{\rm GeV}$.

In order to use the running coupling we must also specify its behaviour below 
the Landau pole ($\Lambda_{QCD}$). For these examples the choice made is to 
freeze the value of the coupling at 500 MeV.

For the two-loop running coupling we have
\begin{equation}
\frac{4\pi}{\alpha_S(Q^2)} + \frac{\beta_1}{\beta_0} \ln \left( \frac{\beta_1 \alpha_S(Q^2)}{4 \pi \beta_0 + \beta_1 \alpha_S(Q^2)} \right) = \beta_0 
\ln \left( \frac{Q^2}{\Lambda^{2}_{QCD}} \right),
\end{equation}
with
\begin{eqnarray}
\beta_0 &=& 11 - \frac{2}{3}n_f, \\
\beta_1 &=& \left( 102 - \frac{38}{3} n_f \right),
\end{eqnarray}
where $n_f$ is the number of flavours; for these examples this is fixed at 5.

This can be solved numerically both for the choice of $\Lambda^2_{QCD}$ and 
for the value of the running coupling. In this case we find $\Lambda_{QCD}$ to 
be $230^{+ 191}_{- 121}~{\rm MeV}$. 
Of course, there are many technical issues related to the running coupling. 
For example matching the running at the threshold of heavy quark masses. For 
simplicity we ignore these issues in the numerical results that proceed.

We now define the variation in the running coupling to be
\begin{eqnarray}
\delta \alpha_S^{+}(Q^2) &=& \alpha_S(Q^2; \Lambda^{2}_{QCD}+
\delta \Lambda^{2+})- \alpha_S(Q^2; \Lambda^2_{QCD})\\
\delta \alpha_S^{-}(Q^2) &=& \alpha_S(Q^2; \Lambda^2_{QCD} - 
\delta \Lambda^{2-}) - \alpha_S(Q^2; \Lambda^2_{QCD}).
\end{eqnarray}
With this we can define the weight for each emission due to the variation of
the running coupling as
\begin{equation}
w_{\alpha_i}^\pm = \left(1 + \frac{\delta \alpha^{\pm}(z_i,\tilde{q}_i^2)}
{\alpha(z_i, \tilde{q}_i^2)} \right) \exp \left( - 
\int_{\tilde{q}_{i}^2}^{\tilde{q}_{i-1}^2} \frac{d \tilde{q}^2}{\tilde{q}^2} 
\int_{z_i^-(\tilde{q}^2)}^{z_i^+(\tilde{q}^2)} dz \, \delta 
\alpha^\pm_S(z, \tilde{q}^2) P_{qq}(z,\tilde{q}^2) \right).
\end{equation}
This weight is normalized to the unvaried weight 1 MC event.

\begin{figure}[t]
\centering
\epsfig{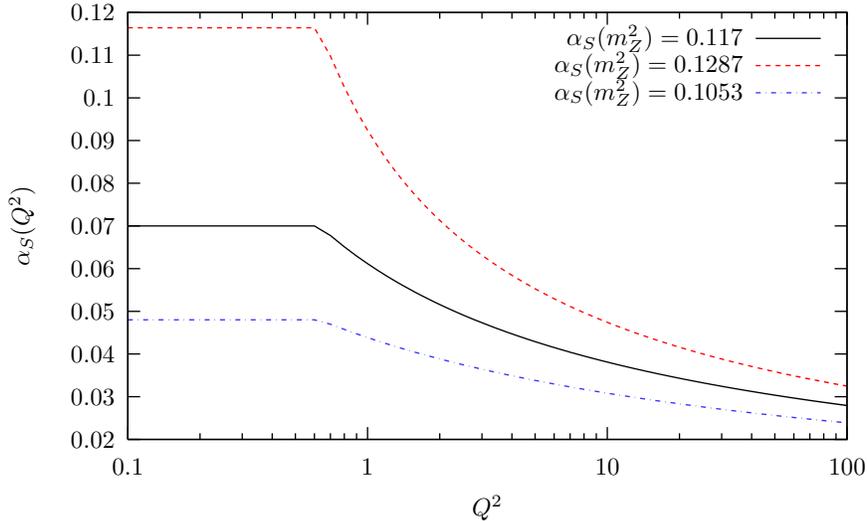}
\caption{The value of the running coupling and the bounds given by the input
at $m_Z^2$.\label{fig:running}}
\end{figure}

Figure~\ref{fig:running} shows the range of values for the running coupling
for the two-loop RGE with a 10\% variation in the value at $m_Z^2$. We can see 
that as one approaches $\Lambda^2_{QCD}$ the variation gets larger. This 
region with a small scale is often where the emissions from the parton 
shower reside; therefore the uncertainty in the measurement of the
running coupling can have a non-neglible effect on the MC predictions.

\begin{figure}
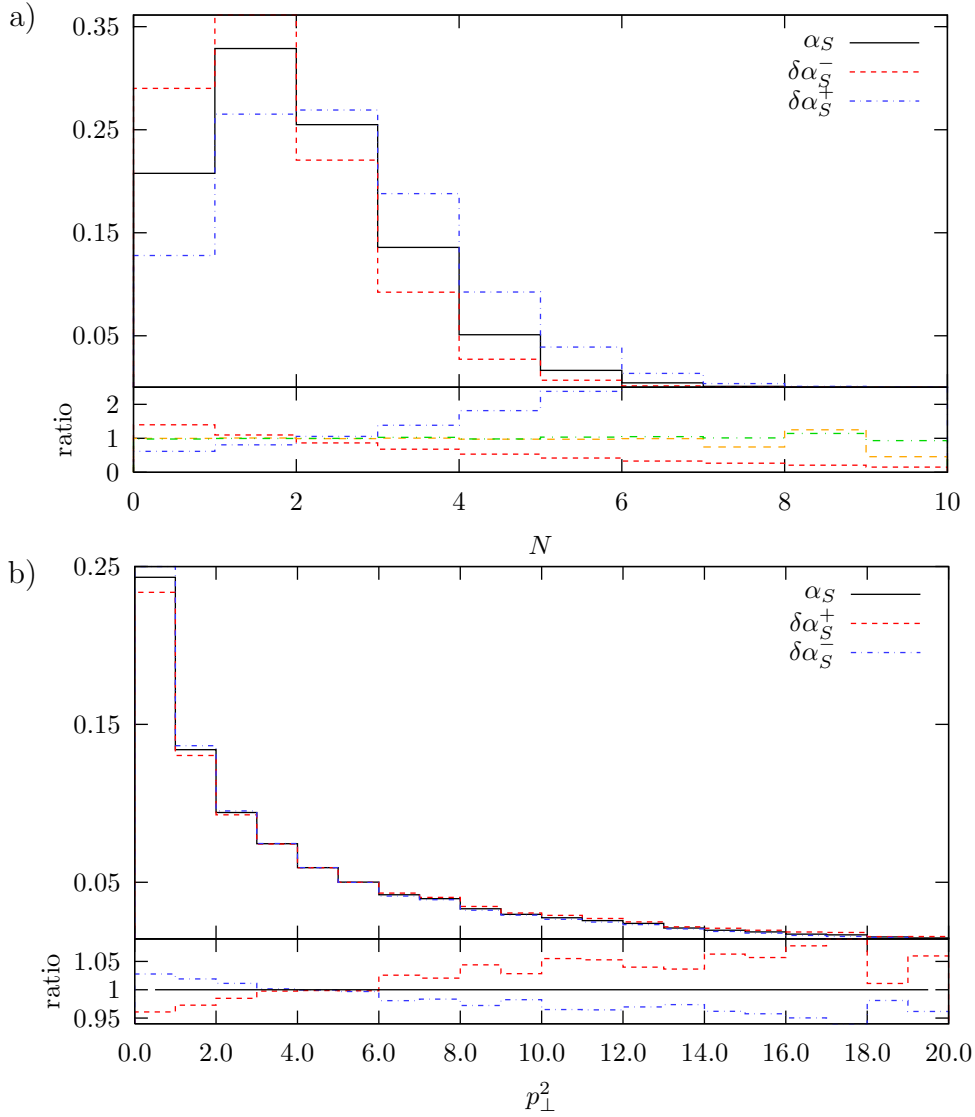

\centering
\epsfig{file=plots.1}
\put(-360,200){a)}\\
\epsfig{file=plots.3}
\put(-365,200){b)}
\caption{a) The distribution of the number of emissions for $m^2 = 0$ and
$\tilde{q}^2 = (100~{\rm GeV})^2$. b) The $p_\perp^2$
distribution. In both cases the solid line is the unvaried case while the
dashed in the upper bound and the dot-dashed is the lower bound. Additionally,
both figures contain a second panel which shows the ratio of the varied to 
unvaried results. In a) we have also included the
ratio of the reweighted MC vs.\ an alternate MC sample generated by changing the
central value of the running coupling to the upper or lower bound of the 
variance. \label{fig:runProb}}
\end{figure}

Figure~\ref{fig:runProb} shows the effect of the variations on a) the number
of emissions and b) the $p_{\perp}^2$-spectrum. Each of these plots is divided
into two panels. The top panel shows the results while the bottom panel shows
the ratio of the reweighted vs.\ unweighted MC. Here we see the behaviour one 
would expect. The higher bound of the coupling gives more emissions, and a
smaller $p_\perp^2$-spectrum while the lower bound does the opposite. In this
figure we chose a massless quark with initial scale $\tilde{q}^2 = (1~ 
{\rm TeV})^2$. 

In figure~\ref{fig:runProb}a we have also included in the ratio panel the
ratio of the reweighted MC vs.\ an alternate MC sample generated by changing the
central value of the running coupling to the upper or lower bound of the 
variance. We see from these results that the reweighted shower does produce 
the same results as reimplementing the shower with the changed running 
coupling.

\section{Kernel Variations}
What may be of more interest from a theoretical side is to vary the structure
of the splitting kernel itself. For example, one could start with the collinear
splitting kernels and vary them by the mass dependent quasi-collinear kernels
to see whether such changes introduce dramatic effects on a set of observables.

The benefit to the procedure presented here is that there is no need to change
the fundamental structure of a given MC. In fact one could add an option to
their code to keep track of the alternate weights, without changing at all 
their basic MC program logics and structures. One caveat is that though this 
method will give an accurate 
estimate of the variations given, this is only true for regions of phase space
in which the original MC fills. If some regions of phase space are empty, or
rarely entered, the changes in that region due to the variation will still 
lack significant statistics. 

For completeness, we present a study which shows the effect of introducing the
quasi-collinear kernel as a variation on the collinear kernel. The collinear 
kernel is simply
\begin{equation}
P_{qq}(z) = \frac{1+z^2}{1-z}.
\end{equation}

To obtain the quasi-collinear kernel, eqn.~(\ref{eqn:kernel}), we must
define a variance of
\begin{equation}
\delta P_{qq}(z,\tilde{q}^2) = -\frac{2 m^2}{z(1-z) \tilde{q}^2}.\label{eqn:vark}
\end{equation}
With this variance we find the alternate weight, for the $i$th emission, 
is given by
\begin{equation}
w_{P_i} = \left( 1 + \frac{\delta P_{qq}(z_i,\tilde{q_i}^2)}
{P_{qq}(z_i, \tilde{q_i}^2)} \right) \exp \left( 
- \int_{\tilde{q}_i^2}^{\tilde{q}_{i-1}^2} 
\frac{d \tilde{q}^2}{\tilde{q}^2} \int_{z^-_i}^{z^+_i} d z \, \alpha_S\left[ 
z^2(1-z)^2 \tilde{q}^2\right] \delta P_{qq}(z, \tilde{q}^2) \right),
\end{equation}
and the total weight due to the kernel variation is the product of the
weight for each emission. This weight is normalized to a weight 1 event with
no variations.

\begin{figure}
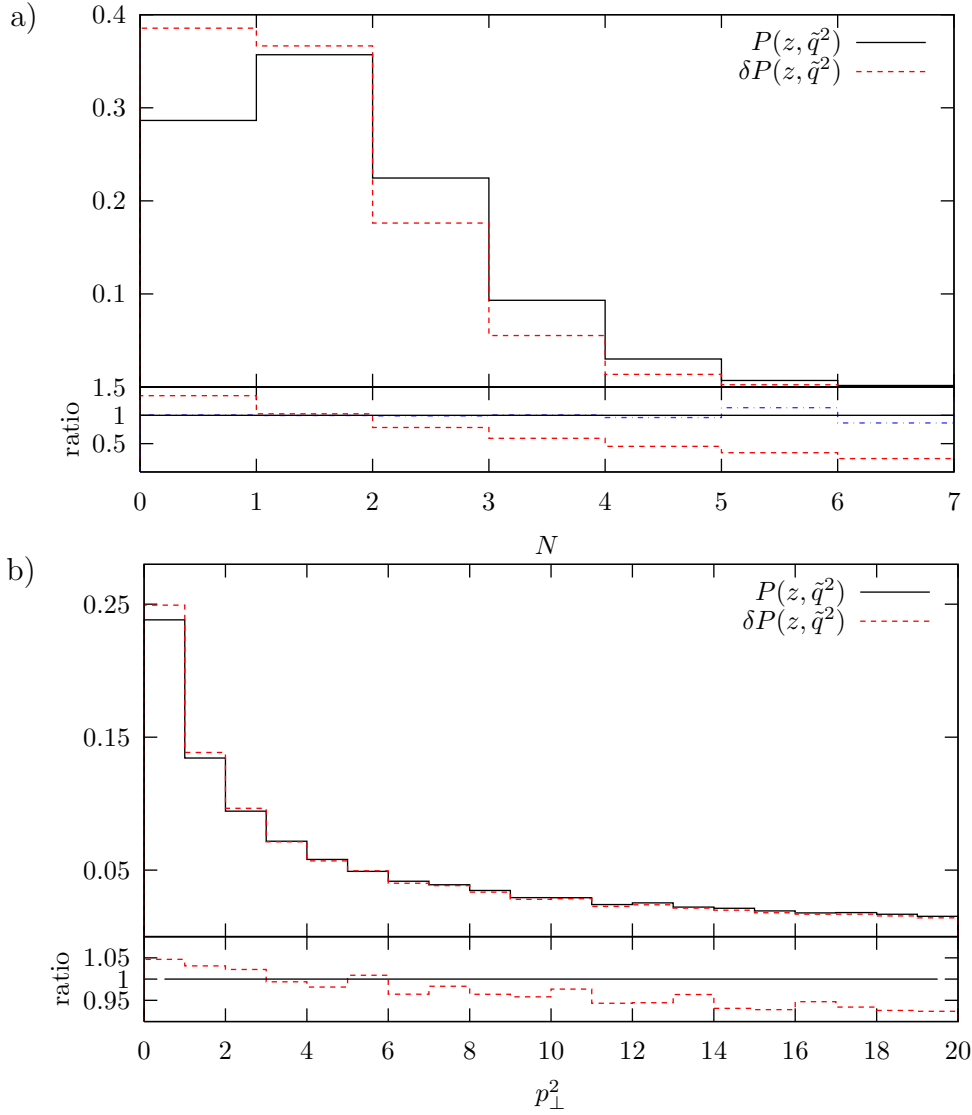

\centering
\epsfig{file=plots.4}
\put(-360,200){a)}\\
\epsfig{file=plots.6}
\put(-365,200){b)}
\caption{a) The distribution of the number of emissions for the collinear
kernel and the quasi-collinear kernel for $m^2 = (175~{\rm GeV})^2$ and
$\tilde{q}^2 = (1 {\rm TeV})^2$. b) The distribution of the $p_\perp^2$
of the outgoing quark for the collinear and quasi-collinear cases under the 
same conditions as a). The solid line shows the
result when the quasi-collinear kernel is used, the dashed line shows the 
result when the variation in eqn.~(\ref{eqn:vark}) is applied and the events 
are weighted. Again, the second panel shows the ratio of the varied to the 
unvaried MC. In figure a) also included is the ratio of the 
reweighted MC vs.\ an alternate MC sample created by changing the kernel in
the MC to the quasi-collinear kernel.
\label{fig:varkern}}
\end{figure}

We now show the result of this variation in the fig.~\ref{fig:varkern} when
showering a top quark with mass $175~{\rm GeV}$ from an initial scale of
$1~{\rm Tev}$. In 
fig.~\ref{fig:varkern}a we show the effect that the quasi-collinear variation
has on the distribution of the number of emissions. As would be expected, for 
larger masses we have fewer emissions. Fig.~\ref{fig:varkern}b 
shows the $p_\perp^2$ spectrum of the outgoing quark. As before these figures
are divided into two panels. The top panel shows the results while the bottom
panel shows the ratio of the reweighted MC vs.\ the unweighted one. Again, in
fig.~\ref{fig:varkern}a the ratio panel also includes the ratio of the 
reweighted MC vs.\ an alternate MC sample created by changing the kernel in
the MC to the quasi-collinear kernel. Again we see that this ratio is 
1 with small variations.

Though we do not provide any numerical results of varying both the kernel and
the running coupling simultaneously, we will present the formulae for these 
weights. In the time-like evolution that we have
discussed, we can define the weight for emission $i$ from 
eqn.~(\ref{eqn:mweight}) as
\begin{eqnarray}
w_i^\pm &=& \left( 1 + \frac{\delta \alpha_S^\pm(z_i,\tilde{q}^2_i)}
{\alpha_S(z_i,\tilde{q}^2_i)} + \frac{\delta P_{qq}(z_i,\tilde{q}^2_i)}
{P_{qq}(z_i,\tilde{q}^2_i)}\right) \label{eqn:wboth} \\
&& \times \exp \left[ - \int^{\tilde{q}^2_{i-1}}_{\tilde{q}^2_i} 
\frac{d \tilde{q}^2} {\tilde{q}^2} \int_{z^-(\tilde{q}^2)}^{z^+(\tilde{q}^2)} 
d z \left( P_{qq}(\cdot) \delta \alpha_S^\pm(\cdot) + 
\delta P_{qq} (\cdot) \alpha_S(\cdot) \right) \right], \nonumber
\end{eqnarray}
and the weight for the event by $w_E = \prod_i w_i^\pm$. In 
eqn.~(\ref{eqn:wboth}) the $\cdot$ represents the pair $(z,\tilde{q}^2)$.
\subsection{Combining Kernel with Running Coupling}
Another potential variation that may be of interest is to vary the kernel
by a term proportional to the running coupling. Such a variation could be used 
to introduce some NLO effects into the kernel.

We must now determine the appropriate variations in this case. We start with 
the general case for the sum of terms up to $\alpha_S^N$
\begin{equation}
F[(\alpha_S,P^{(1)}_{qq},...,P^{(N)}_{qq})(z,\tilde{q}^2)]= 
\sum_{i=1}^N \left( \frac{\alpha_S(z,\tilde{q}^2)}{2 \pi}\right)^i 
P^{(i)}_{qq}(z, \tilde{q}^2),
\end{equation}
of which we now have $N+1$ functions to vary over. We compute
$\delta F[{\bphi}] = F[{\bphi} + \delta {\bphi}] - F[{\bphi}]$ and keep the 
lowest order in the variations.
\begin{eqnarray}
\delta F[{\bphi}] &=& \sum_i \left( \alpha_S+\delta \alpha_S \right)^i 
\left(P^{(i)} + \delta P^{(i)} \right) - \sum_i \alpha_S^i P^{(i)}, 
\nonumber \\
&\approx& \sum_i \left( i \alpha_S^{i-1} \delta \alpha_S P^{(i)} +
\alpha_S^i \delta P^{(i)} \right).
\end{eqnarray}
which is equivalent to the functional derivative. If we keep higher order terms
the equation is
\begin{equation}
\delta F = \sum_{i=1}^N \left[ \sum_{j=1}^i \left( \begin{array}{c} j \\ i 
\end{array} \right) \alpha_S^{i-j} \delta \alpha_S^j \left( P^{(i)}_{qq} +
\delta P^{(i)}_{qq} \right) + \alpha_S^i \delta P^{(i)}_{qq} \right],
\end{equation}
where $N$ is the length of the vector ${\bm\vhi}$.

For the examples given here we will look at only $N=2$ and set $P^{(2)} = 0$
and study the variation around that choice. This means we have
\begin{equation}
\delta F = \delta \alpha_S \left(P_{qq}^{(1)} + \delta P_{qq}^{(1)} 
\right) + \left[ 2 \alpha_S \delta \alpha_S + \delta \alpha_S^2 \right] 
\delta P_{qq}^{(2)} + \alpha_S \delta 
P^{(1)}_{qq} + \alpha_S^2 \delta P^{(2)}_{qq}.
\end{equation}
If we consider only the lowest order in the variations
\begin{equation}
\delta F \approx \delta \alpha_S P_{qq}^{(1)} + \alpha_S \delta 
P^{(1)}_{qq} + \alpha_S^2 \delta P^{(2)}_{qq}.
\end{equation}

We now turn to an example. We choose the form of $\delta P^{(2)}_{qq}(z)$ 
according to full NLO kernel~\cite{Furmanski:1980cm,Curci:1980uw}. This is composed of two parts, the 
flavour singlet (S) and non-singlet (V) contributions
\begin{equation}
\delta P^{(2)}_{qq}(z,\tilde{q}^2) = P^{S(2)}_{qq}(z) + P^{V(2)}_{qq}(z),
\end{equation}
where these functions are defined in the appendix. We choose 
$\delta P^{(1)} = 0$ and $\delta \alpha = 0$ for these examples.

\begin{figure}
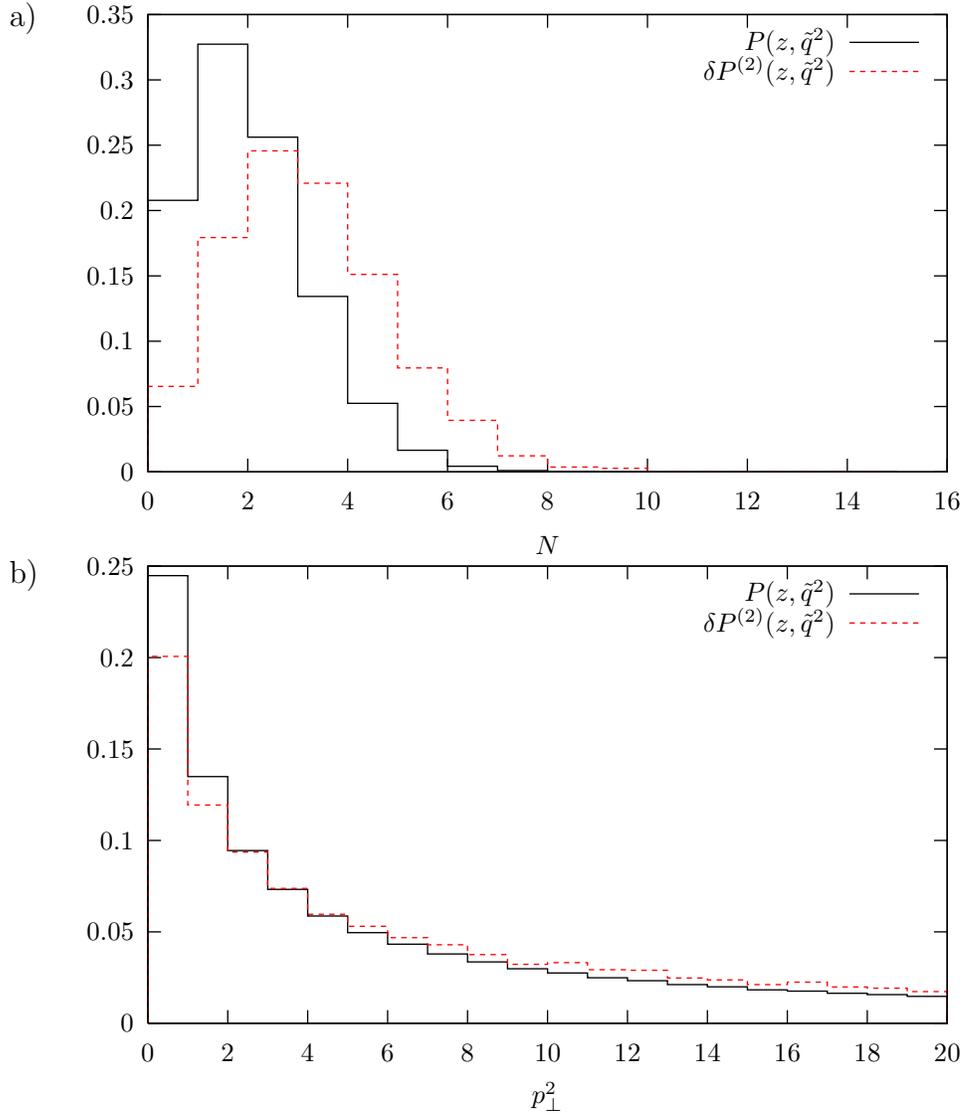

\centering
\epsfig{file=plots.7}
\put(-360,200){a)}\\
\epsfig{file=plots.8}
\put(-360,200){b)}
\caption{a) The distribution of the number of emissions using the collinear
kernel at ${\cal O}(\alpha_S)$ and applying the variation discusses in the text
at ${\cal O}(\alpha_S^2)$. b) The $p_\perp^2$ distribution of the outgoing
quark under the same conditions as a). \label{fig:NLO}}
\end{figure}

Figure~\ref{fig:NLO}a shows the effect on the number of emissions and 
fig.~\ref{fig:NLO}b shows the effect on the $p_\perp^2$--spectrum of the
outgoing quark line. We see that the number of emissions is slightly higher
with a harder spectrum.

The construction of a next-to-leading log (NLL) parton shower has the problem 
of negative values for the splitting kernels. These destroy the 
probabilistic interpretation of the Sudakov form factors. Naively, one would 
assume that this will destroy any meaningful results for the NLL weights. In
our case, this is not true. We are reweighting the total density according
to the NLL corrections. These may introduce large or negative weights to the
reweighted shower, but this is necessary as this correctly describes the
density. In the inclusive picture, these negative weights are integrated over
and pose no problem; exclusively, these negative weights must be treated
correctly in the analysis.

\section{Variation of Kinematics}
We now consider another use of the alternate weights. Here we
wish to use these weights to transform one parton shower into another. This,
of course, is not an exact transformation. This requires additional 
knowledge about the structure of the alternate parton shower.

The idea is to use the variables generated by one shower and reshape the 
distribution to give the results if an alternate shower was used. In this
section we discuss the intrinsic kinematical definitions. 

Consider a new kinematics, similar to the one used in 
Pythia~\cite{Sjostrand:2000wi}. Here
we wish to order the parton shower in virtuality ($Q^2$). This requires a 
mapping from $\tilde{q}^2$ into $Q^2$. First we have the definition of the 
transverse momentum as
\begin{equation}
{\bm p}_\perp^2 = \bar{z}(1-\bar{z})(Q^2-Q_0^2),
\end{equation}
where $\bar{z}$ is the momentum fraction in the Pythia-like kinematics.
From eqn.~(\ref{eqn:qtilde}) we find (neglecting $Q_0^2$)
\begin{equation}
Q^2 = \frac{z^2(1-z)^2\tilde{q}^2 - \mu^2(1-z)^2 - z Q_g^2}{\bar{z}(1-\bar{z})}.
\label{eqn:qtoQ}
\end{equation}
We also have the boundary of real phase space given by the requirement that
there is a real $p_{\perp}^2$ and the imposition of a particular ordering 
scheme. 

There is a different interpretation of the meaning of the momentum
fraction $z$ in the Pythia-like and Herwig-like shower; they have the same 
distribution, however. We compensate for this by constructing the full four-momentum 
from the Herwig-like shower and deconstructing 
the associated variables for each emission. The weights can then be computed 
from this. This method has the additional benefit that the four momentum 
configuration is identical in both cases; thus hadronization effects and hadron
decays are identical.

We now turn to the structure of the probability density itself. For our
original kinematics we find (for the massless case)
\begin{equation}
F[(\alpha_S, P_{qq})(z,\tilde{q}^2)] = \frac{\alpha_S(z^2(1-z)^2 \tilde{q}^2)}
{2\pi} \frac{P_{qq}(z)}{\tilde{q}^2} \theta(\tilde{q}^2_{i-1}-\tilde{q}^2)
\theta(z^+-z) \theta(z-z^-),
\end{equation}
while in the Pythia-like shower we have
\begin{equation}
\bar{F}[(\alpha_S, P_{qq})(\bar{z},Q^2)] = \frac{\alpha_S(\bar{z}(1-\bar{z})Q^2)}{2\pi}
\frac{P_{qq}(\bar{z})}{Q^2} \theta(Q^2_{i-1}-Q^2) \theta(\bar{z}^+-\bar{z}) 
\theta(\bar{z}-\bar{z}^-).
\end{equation}
From these we can define our variations such that 
\begin{equation}
\bar{F}[(\alpha_S, P_{qq})(\bar{z},Q^2)] = F[(\alpha_S, P_{qq})(z,\tilde{q}^2)] + 
\delta F,
\end{equation}
thus
\begin{equation}
\bar{\cal P}(\bar{z},Q^2) d Q^2 d \bar{z} = ({\cal P} + \delta {\cal P})(z,\tilde{q}^2) 
d \tilde{q}^2 d z.
\end{equation}
From this we find
\begin{equation}
\delta F = \bar{F}[(\alpha_S, P_{qq})({\cal T}(z,\tilde{q}^2))] {\cal J}
(\bar{z},Q^2) - F[(\alpha_S, P_{qq})(z, \tilde{q}^2)],
\end{equation}
where ${\cal J}$ is the Jacobian factor for the coordinate transformation
${\cal T}(z, \tilde{q}^2)$. These are defined in the appendix. At this point we 
can exploit the analytic structure
of the Sudakov form factor,
\begin{equation}
\Delta(t;t_0) = \Delta(t;t_1) \Delta(t_1;t_0).
\end{equation}
This allows use to seperate the weights into the real and the Sudakov
components and to calculate the Sudakov components over the full evolution
scale, rather than just the scales between each emission. This gives
\begin{equation}
w_\Delta = \frac{\Delta_P(Q^2_{ini} ; Q_0^2)}{\Delta_H(\tilde{q}^2_{ini},
\tilde{q}^2_0)}.
\end{equation}
The resulting weight for the real emissions is
\begin{equation}
w^{(R)}_i = \frac{\alpha_S(\bar{z}_i(1-\bar{z}_i)Q_i^2) P_{qq}^{(P)}(\bar{z}_i) 
\theta(Q_{i-1}^2 - Q_i^2) \theta(\bar{z}^+ - \bar{z}) \theta(\bar{z}-\bar{z}^-)}
{\alpha_S(z_i^2(1-z_i)^2  \tilde{q}^2_i) P_{qq}^{(H)}(z,\tilde{q}^2)} \frac{\tilde{q}^2 
{\cal J}(\bar{z}, Q^2)}{Q^2}.
\end{equation}
Here the $\theta$ functions for the Herwig-like evolution are ignored as they
are always fulfilled by the original shower construction. The total weight is
given simply as
\begin{equation}
w = w_\Delta \prod_{i=1}^N w_i^{(R)}.
\end{equation}

The question now is what does the weighted shower physically give us? This 
gives us the weight, relative to the unweighted original shower, of producing
the kinematical configuration via the other shower. For our example here this
means that it will weight our Herwig-like shower to be that of the Pythia-like
construction. Our weighted shower will produce events that are both ordered in
virtuality and in angle. Comparing the weighted results versus an independent
implementation of the full Pythia-like shower would illustrate, for any 
observable, the effect of the different limits in phase-space inherent in each
implementation. Furthermore, it could be used to illustrate the effects
of alternate choices of ordering; e.g.\ colour connections between jets.

To illustrate this technique we use as a model $e+e-$ annihilation into
a $q \bar{q}$ pair. This pair then undergoes final state radiation, but
the subsequent emissions do not. We reconstruct the kinematics of the event
and, in order to conserve $\sqrt{s}$, we rescale each jet by a common factor,
$k$, such that
\begin{equation}
\sqrt{s} = \sum_{i=1}^N \sqrt{q_i^2 + k {\bm p}_i^2},
\end{equation}
where $q_i^2$ is the virtuality of jet $i$. To illustrate the reweighting
between the Herwig-like and Pythia-like shower we study the thrust observable.
This is given by
\begin{equation}
T = \max_{\bm n} \frac{\sum_{i=1}^N \left| {\bm p}_i \cdot {\bm n} \right|}
{\sum_{i=1}^N \left| {\bm p}_i \right|}.
\end{equation}
This observable was chosen as the thrust has a strong correlation to the
hardest emission, but also is effected by subsequent emissions. As we do not 
shower the emitted gluons, studying an observable which have a strong 
dependence on 2 or more emissions is not as illustrative.

\begin{figure}
\centering
\epsfig{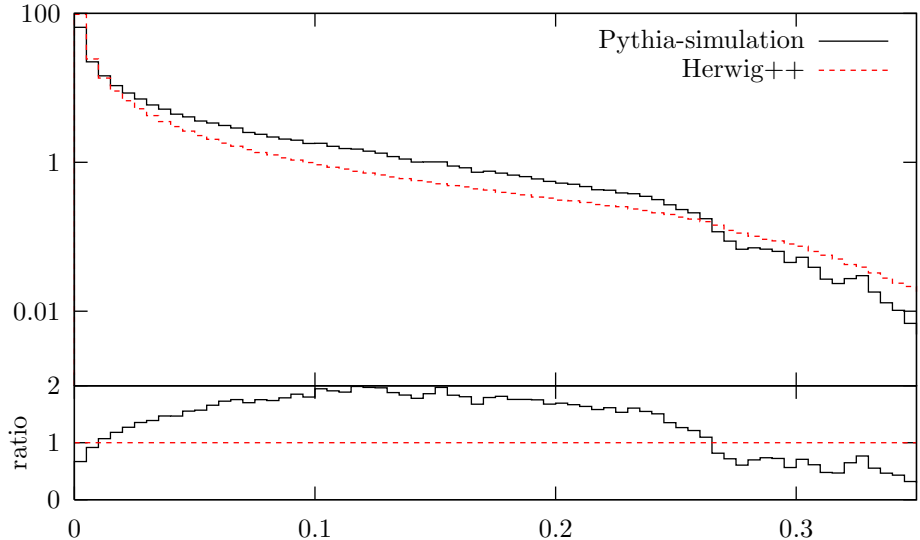}
\caption{$1-T$ for the Herwig-like shower and reweighted to a Pythia-like
shower, as described in the text. These differences are due to the different
kinematics definitions used in each shower. The bottom panel shows the ratio of
the Pythia-like vs. Herwig-like.\label{fig:thrust}}
\end{figure}

Figure~\ref{fig:thrust} shows the result for $\sqrt{s} = 1~{\rm TeV}$. We can see
the deviations, and as expected they are not too large. As these are not
the result of a full event generation it is not useful to compare these to 
data.

\section{Conclusion}
We have presented a new approach to understanding the errors associated with a
MC prediction. This approach can be added to almost all currently existing MC 
programs without changing the physics or the behaviour of the code. Instead, 
we have provided a method to track alternate weights for events. These 
alternate weights provide the tool to reshape MC predictions to see what such 
a prediction would be if various pieces of the MC were altered.

Though this technique is quite successful, it cannot compensate for all 
possible alterations. As this algorithm provides an alternate weight for an 
event generated by a MC it cannot provide events which cannot be generated by 
the original MC. This means that some of the physical limitations of an 
already existing code cannot be overcome through this method. We do not see 
this as a drawback, however. The purpose of this technique is to understand 
the physics and the limitations inherent in a MC implementation. To this end, 
such limitations of this technique can provide valuable insight.

This paper has provided numerical examples of a toy parton shower model based 
on the real MC behaviour of Herwig++~\cite{Gieseke:2003hm,Gieseke:2003rz}. It 
may be quite 
illustrative to apply this method to a fully featured general purpose MC, 
including hadronization and hadron decay, to see how much variation exsists in 
such a parton shower implementation. With such an implementation one may be able
to check the accuracy of many MC predictions and to understand the limitations 
of these predictions.

\subsubsection*{Acknowledgment}
The authors would like to thank S. Jadach and Z. Was for many useful discussions.

\begin{appendix}
\section{NLO splitting function}
Here we present the formulae for the NLO splitting functions used in this
paper. These are defined in the $\overline{\rm MS}$ factorization/renormalization
scheme. First we present the flavour singlet contribution~\cite{Furmanski:1980cm}
\begin{eqnarray}
P^{S(2)}_{qq}(z) &=& C_F^2 \left\{ -1 + z  +\frac{1}{2} \left[(1-3z) \ln z
- (1+z) \ln^2 z \right] \right. \nonumber \\
&& \left. - \left[ \frac{3}{2} \ln z + 2 \ln z \ln (1-z) \right] p_{qq}(z) + 
2 p_{qq}(-z) S_2(z) \right\} \nonumber \\
&& + C_F C_A \left\{ \frac{14}{3}(1-z) + \left[ \frac{11}{6} \ln z + 
\frac{1}{2}\ln^2 z + \frac{67}{18} - \frac{\pi^2}{6} \right] p_{qq}(z) \right.
\nonumber \\
&& \left. -p_{qq}(-z) S_2(z) \right\} \nonumber \\
&& C_F T_f \left\{ -\frac{16}{3} + \frac{40}{3} z + \left( 10 z + \frac{16}{3} 
z^2 + 2 \right) \ln z \right. \nonumber \\
&& \left. -\frac{112}{9} z^2 + \frac{40}{9z} - 2 (1+z) \ln^2 z - \left[ 
\frac{10}{9} + \frac{2}{3} \ln z \right] p_{qq}(z) \right\}.
\end{eqnarray}
In this formula we have
\begin{eqnarray}
p_{qq}(z) &=& \frac{2}{1-z} - 1 - z, \\
S_2(z) &=& \int_{\frac{z}{1+z}}^{\frac{1}{1+z}} \frac{d z}{z} \ln \left(
\frac{1-z}{z} \right), \nonumber \\
&=& -2 {\rm Li}_2(-z) + \frac{1}{2} \ln^2 z - 2 \ln z \ln (1+z) - \frac{\pi^2}
{6}, \\
C_F &=& \frac{4}{3}\, ; \,  C_A = 3\, ; \, T_f = \frac{1}{2} n_f.
\end{eqnarray}

Now we give the flavour non-singlet function~\cite{Curci:1980uw}
\begin{eqnarray}
P^{V(2)}_{qq} &=& C_F^2 \left\{ - \left[ 2 \ln z \ln (1-z) + \frac{3}{2} \ln z
\right]p_{qq}(z) \right. \nonumber \\
&& \left.  - \left( \frac{3}{2} + \frac{7}{2} z\right) \ln z - \frac{1}{2} 
(1+z) \ln^2 z - 5(1-z) \right\} \nonumber \\
&& C_F C_A \left\{ \left[ \frac{1}{2} \ln^2 z + \frac{11}{6} \ln z + 
\frac{67}{18} - \frac{\pi^2}{6} \right] p_{qq}(z) \right. \nonumber \\
&& \left. (1+z) \ln z + \frac{20}{3}(1-z) \right\} \nonumber \\
&& + C_F T_f \left\{ - \left[ \frac{2}{3} \ln z + \frac{10}{9} \right] 
p_{qq}(z) - \frac{4}{3} \left(1-z\right) \right\}.
\end{eqnarray}

We note that the superscripts given here differ from the normal convention. 
Here they indicate the total number of powers of $\alpha_S$ in the branching
probability. Normal convention decrements these by one to indicate the
total order of expansion in the splitting kernel.

\section{Coordinate Transformation Jacobian}
In this appendix we compute the full Jacobian factor for the transformation between the
evolution variables used in the Herwig-like shower and those used in the Pythia-like
shower. This transformation has the form
\begin{equation}
{\cal T}(\tilde{q}^2, z) \to (Q^2, \bar{z}).
\end{equation}
This is to be done after the full momentum reconstruction so we know all
of the components of the momenta. We need to numerically evaluate the 
Jacobian factor for the weights of the real emissions.

We start with
\begin{equation}
\tilde{q}^2 = \frac{\vec{p}_\perp^2}{z^2 (1-z)^2} + \frac{\mu^2}{z^2}
+ \frac{Q_g^2}{z(1-z)^2},
\end{equation}
and 
\begin{equation}
Q_i^2 = q_{i-1}^2 = \frac{q_i^2}{z} + \frac{k_i^2}{1-z} +
\frac{\vec{p}_\perp^2}{z (1-z)}.
\end{equation}
Together we find
\begin{equation}
Q_i^2 = \frac{q_i^2 + (z-1)\mu^2}{z} + \tilde{q}^2 z(1-z).
\end{equation} 

We now find $\bar{z}$ as
\begin{equation}
\bar{z} = \frac{q_i^0}{q_{i-1}^0} = \frac{\alpha_i p^0 + \beta_i n^0}{q_{i-1}^0}
= \frac{z \alpha_{i-1} p^0 + \beta_i n^0}{q_{i-1}^0}.
\end{equation}
From the Herwig++ variables we have
\begin{eqnarray}
\beta_i &=& \frac{q_i^2 - q_\perp^2 - \alpha_i^2 m^2}{2 \alpha_i p \cdot n} \\
&=& \frac{q_i^2 - q_\perp^2 - z^2 \alpha_{i-1}^2 m^2}{2 z \alpha_{i-1} 
p \cdot n}.
\end{eqnarray}
We now use
\begin{eqnarray}
\vec{q}_{\perp i} &=& \vec{p}_{\perp i} - z \vec{q}_{\perp i-1}, \\
\gamma &=& 2 \alpha_{i-1} p \cdot n,
\end{eqnarray}
to find
\begin{eqnarray}
\beta_i &=& \frac{1}{\gamma z} \left[q_i^2 + \tilde{q}^2 z^2 (1-z)^2
- \mu^2 (1-z)^2 + z Q_g^2 \right. \nonumber \\
&& + 2 z \sqrt{\tilde{q}^2 z^2 (1-z)^2
- \mu^2 (1-z)^2 + z Q_g^2} \, |\vec{q}_{\perp i-1}| \cos \theta 
\nonumber \\
&& \left. + \vec{q}_{\perp i-1}^2 - z_i^2 \alpha_{i-1}^2 m^2 \right].
\end{eqnarray}

Using these formulae we are now able to compute the full Jacobian
\begin{equation}
{\cal J}(\bar{z},Q^2) = \frac{\partial (Q^2, \bar{z})}{\partial (\tilde{q}^2, z)}
= \left| \begin{array}{cc} \frac{\partial Q^2}{\partial \tilde{q}^2} &
\frac{\partial \bar{z}}{\partial \tilde{q}^2} \\ \frac{\partial Q^2}{\partial
z} & \frac{\partial \bar{z}}{\partial z} \end{array} \right|.
\end{equation}

We have
\begin{eqnarray}
\frac{\partial Q^2}{\partial \tilde{q}^2} &=& z (1-z), \\
\frac{\partial Q^2}{\partial z} &=& \frac{\mu^2 - q_i^2}{z^2} + \tilde{q}^2
(1 - 2 z), \\
\frac{\partial \bar{z}}{\partial \tilde{q}^2} &=& \frac{\beta_i}{\partial 
\tilde{q}^2} \frac{n^0}{q_{i-1}^0}, \\
\frac{\partial \bar{z}}{\partial z} &=& \frac{\alpha_{i-1} p^0}{q_{i-1}^0}
+ \frac{\partial \beta_i}{\partial z} \frac{n^0}{q_{i-1}^0}. \\
\frac{\partial \beta_i}{\partial \tilde{q}^2} &=& z^2 (1-z)^2 
\left( 1 + \frac{z |\vec{q}_{\perp i-1}|\cos \theta}
{\sqrt{\tilde{q}^2 z^2 (1-z)^2
- \mu^2 (1-z)^2 + z Q_g^2}} \right), \\
\frac{\partial \beta_i}{\partial z} &=& -\frac{q_i^2}{\gamma z^2} + 
\frac{\tilde{q}^2 (1-z^2)}{\gamma} - \frac{\mu^2}{\gamma} \left( 
\frac{z^2-1}{z^2} \right) \\
&& + \frac{\tilde{q}^2 2 z (1-z)(1-2z) + 2 \mu^2 (1-z) + Q_g}
{\gamma \sqrt{\tilde{q}^2 z^2 (1-z)^2 - \mu^2 (1-z)^2 + z Q_g^2}}
|\vec{q}_{\perp i-1}|\cos \theta - \frac{\alpha_{i-1}^2 m^2}{\gamma}. \nonumber
\end{eqnarray}

We can now completely determine the Jacobian factor for transforming between
the two kinematics. Thus the weight for the real emission is
\begin{equation}
w^{(R)}_i = \frac{\alpha_S(\bar{z}(1-\bar{z}) Q^2) P^{(P)}_{qq}(\bar{z},Q^2) {\cal J}
(\bar{z},Q^2)}{\alpha_S(z^2 (1-z^2) \tilde{q}^2) P^{(H)}_{qq}(z, 
\tilde{q}^2)} \frac{\tilde{q}^2}{Q^2}.
\end{equation}

Finally, it is left to compute the cutoff virtuality from the cutoff in 
$\tilde{q}^2$. Since we know the outgoing quark has virtuality $\mu^2$ we
assume that this is the cutoff on the virtuality ordered shower. One could 
find from
\begin{equation}
\tilde{q}^2_0 = \frac{\mu^2}{z^2} + \frac{Q_g^2}{z (1-z)^2}
\end{equation}
the value of $z$ to use in
\begin{equation}
Q_0^2 = \tilde{q}^2_0 z (1-z) + \mu^2.
\end{equation}
The weight for the Sudakov factor is then
\begin{equation}
w_\Delta = \frac{\exp \left(- \int_{Q_0^2}^{Q_{max}^2} \frac{d Q^2}{Q^2} 
\int_{z^-_P(Q^2)}^{z^+_P(Q^2)} d \bar{z} \frac{\alpha_S(\bar{z} (1-\bar{z}) Q^2)}{2\pi}
P^{(P)}_{qq}(\bar{z},Q^2) \right)}
{\exp \left(- \int_{\tilde{q}_0^2}^{\tilde{q}_{max}^2} \frac{d \tilde{q}^2}
{\tilde{q}^2} \int_{z^-_H(\tilde{q}^2)}^{z^+_H(\tilde{q}^2)} d z 
\frac{\alpha_S(z^2 (1-z)^2 \tilde{q}^2)}{2\pi} P^{(H)}_{qq}(z,
\tilde{q}^2) \right)}.
\end{equation}

\end{appendix}

\end{document}